\documentclass{article}

\begin{document}

{\Large{\centerline{\bf A New Perspective on Gravity}}}
{\Large{\centerline{\bf and the Dynamics of Spacetime}}}

\bigskip

\centerline{T.~Padmanabhan}

\medskip
\centerline{IUCAA, 
Post Bag 4, Ganeshkhind, Pune - 411 007, India}
\centerline{Email: nabhan@iucaa.ernet.in}

\date{\today}

\bigskip

\begin{abstract}
The Einstein-Hilbert action has a bulk term and a surface term (which arises from integrating a four divergence). I show that
\textit{one can obtain Einstein's equations from the  surface term alone}. This leads to: (i) a novel, completely self-contained, perspective on
gravity and (ii) a concrete mathematical
framework in which 
the description of spacetime dynamics by Einstein's equations is similar to the description of 
a continuum solid in the thermodynamic limit.
\end{abstract}
\bigskip

In general relativity, one can distinguish between the kinematics (spacetime tells matter how to move) and the dynamics (matter tells spacetime how to curve). The 
geometric description of the 
\textit{kinematics} arises quite elegantly  from the principle of equivalence\cite{mtw}. To obtain the dynamics, which depends on the choice of the action principle, 
 one uses the Einstein-Hilbert Lagrangian  $L_{EH}\propto R$ which has a formal structure $L_{EH}\sim R\sim (\partial g)^2+\partial^2g$. 
If the surface term obtained by integrating $L_{sur}\propto \partial^2g$ is ignored (or, more formally, canceled by an extrinsic curvature term) then the Einstein's equations arise from the variation of the bulk
term $L_{bulk}\propto (\partial g)^2$ which is the non-covariant $\Gamma^2$ Lagrangian. 

On closer inspection, this procedure raises several questions (see the reviews, \cite{tp1}). To begin with, it 
 does not have the  elegance or uniqueness, 
possessed by the geometric description of the kinematics. Second, 
in no other field theory (including Yang-Mills) does 
the symmetries of the theory   lead to a Lagrangian involving second derivatives of the dynamical variables;  it is clearly unusual. Third, the action is a window to the quantum theory and one might suspect most of the difficulties in quantising gravity
might be due to quantising the wrong action functional based on wrong fundamental variables; it is possible that continuum spacetime is like an elastic solid (see eg. \cite{sakharov}) and what we should be quantising is the `atomic structure' of spacetime.
Finally, 
in using $L_{bulk}$ to obtain the dynamics, we are also  assuming tacitly that the gravitational degrees of freedom are the components of the metric and they
reside in the volume $\mathcal{V}$. But recall that,
 around any event, one can choose a local inertial frame so that $L_{bulk}\propto (\partial g)^2$ vanishes since $\partial g$ vanishes. On the other hand, one \textit{cannot} make $L_{sur}\propto\partial^2 g$ part to vanish by any choice of coordinates suggesting that \cite{tpholo} \textit{the true  degrees of freedom of gravity
for a volume $\mathcal{V}$
reside in its boundary $\partial\mathcal{V}$}. (This is most easily seen by evaluating the action in Riemann coordinates in which
the bulk vanishes and only $L_{sur}$ contributes).
 
 This point of view is also strongly supported by the study
of horizon entropy, which shows that the degrees of freedom hidden by a horizon scales as the area and not as the volume. 
If  this view is correct,  it \textit{must }be possible to obtain the dynamics of gravity from an approach which  uses \textit{only} the surface term of the Hilbert action. Indeed,
we will show that suitable variation of the surface term will lead to Einstein's equations and that \textit{ we do not need the bulk term at all !}.
 What is more, we will first obtain the surface term itself  from
general considerations thereby providing
a new, self-contained and holographic approach to gravity.

We begin by noting that, in \textit{any} spacetime, there will exist families of observers (congruence of timelike curves) 
who will have access to only part of the spacetime. 
Let a timelike curve $X^a(t)$, parametrized by the proper time $t$ of the clock
   moving along that curve, be the trajectory of an observer in such a congruence and let $\mathcal{C}(t)$
   be the past light cone for the event $\mathcal{P}[X^a(t)]$ on this trajectory. The union $\mathcal{U}$ of all these past light cones 
   $\{\mathcal{C}(t), {\rm all} \ t \}$ determines whether an observer on the trajectory $X^a(t)$
   can receive information from all events in the spacetime or not. If $\mathcal{U}$ has a nontrivial boundary, there will be regions in the spacetime from which this observer cannot receive signals. 
The boundary of the union of causal pasts of \textit{all} the observers in the congruence --- which is essentially  the boundary of the union of backward
light cones ---  will define a {\it causal} horizon $\mathcal{H}$ for this congruence. 
The well known examples are observers at $r=$ constant$>2M$
   in the Schwarzschild spacetime or the uniformly accelerated observers in flat spacetime.
This causal  horizon is
  \textit{dependent} on the family of observers that is chosen, but is \textit{coordinate independent}.

Any class of observers, of course, has an equal right to describe physical phenomena \textit{entirely in terms of the variables defined in the regions accessible to them}. 
The action
functional describing gravity, used by these observers (who have access to only part of the spacetime)  will depend on variables defined on the region accessible to them, including the boundary of this region \cite{tpapoorva}. Since the horizon (and associated boundaries) may exist for some observers (e.g., uniformly accelerated observers in flat spacetime, $r=$ constant $>2M$ observers in the Schwarzschild spacetime ...) but not for others (e.g, inertial observers in flat spacetime, freely falling observers
inside the event horizon, $r<2M$, in the Schwarzschild spacetime ... ), this brings up a new level of observer dependence in the action functional describing the theory. It must, however, be stressed that this view point \textit{is completely in concordance with what we do in other branches of physics, while defining action functionals}. The action describing QED at $10$ MeV, say,
does not use degrees of freedom relevant at $10^{19}$ GeV which we have no access to. Similarly, if an observer has no access to part of the spacetime, (s)he should be able to use an action principle
using the variables (s)he can access, which is essentially the philosophy of renormalisation group theory translated 
from momentum space  into real space. The physics of the region blocked by the horizon will be encoded in a boundary term in the action.
We shall now determine this boundary term.

Since we would like the action to be an integral over a local density, the surface term must arise from integrating a four-divergence term in the Lagrangian and such a surface term (in the \textit{Euclidean} sector, which we shall consider first) will have a generic form:
\begin{equation}
A_{\rm sur}  = \int_{\cal V} d^4 x \sqrt{g} \,\nabla_a U^a 
\label{firsteqn}
\end{equation}
The vector $U^a$ has to be built out of: (i) the normal $u^i$ to the boundary $\partial{\cal V}$ of the region ${\cal V}$, (ii) the metric $g_{ab}$ and (iii)
the covariant derivative operator $\nabla_j$ acting at most  once. (The last restriction arises because the equations of motion
should be of no order higher than two.) The normal $u^i$ is defined only on the boundary $\partial\mathcal{V}$ but we can extend it to the bulk $\mathcal{V}$, forming a vector field, in any manner we like, since the $A_{\rm sur}$ only depends on its value on the boundary.
Allowing the action to depend on $u^i$ (in addition to $g_{ab}$) introduces a
foliation (observer) dependence, though $A_{\rm sur}$ is still generally covariant. A \textit{non trivial}
consistency requirement on our approach is that the dynamical equations which we finally obtain should be independent of $u^i$;
we will see that this is indeed fulfilled.

 Given these conditions,
there are only four possible choices for $U^i$, viz.
$(u^j\nabla^i u_j, u^j\nabla_j u^i, u^i\nabla_j u^j,u^i)$. Of these four, the first one identically vanishes since $u^j$ has constant norm; the second one --- which is
the acceleration $a^i=u^j\nabla_j u^i$ --- gives zero on integration since  the dot product with the normal on the boundary vanishes: $u_iU^i=a^iu_i=0$. Hence the most general vector $U^i$ we need to consider is the linear combination of $u^i$ and $Ku^i$ where $K\equiv -\nabla_iu^i$ is the trace of the extrinsic curvature of the boundary. Of these two,  $U^i=u^i$ will lead to the volume of the bounding surface which we can  ignore. (It merely adds a constant to $K$ and retaining it does not alter any of our conclusions below). Thus the  surface term (arising from $Ku^i$) must have the form 
\begin{equation}
\label{threedaction}
A_{\rm sur}\propto\int_\mathcal{V} d^4 x \, \sqrt{g} \nabla_i (Ku^i)
=
\frac{1}{8\pi G}\int_{\partial\mathcal{V}} d^3 x \, \sqrt{h} K 
\end{equation}
where $G$ is a  constant to be determined (which has the dimensions of area in natural units
with $c=\hbar=1$) and  $8\pi$ factor is introduced with some hindsight. The form of $A_{\rm sur}$, of course, is familiar but
we have determined it  from 
general considerations and \textit{not} through the Einstein-Hilbert action. 

 More importantly, $(-A_{sur})$ has the physical interpretation of the entropy attributed to the horizon\cite{tp1} by these observers. Working in the Euclidean sector,
 near any static horizon one can set up the Rindler coordinates  which has the Euclidean extension (with $\tau=it$): 
\begin{equation}
ds^2_E\approx N^2 d\tau^2 +dN^2/\kappa^2+dL_\perp^2
\label{eext}
\end{equation} 
where $\kappa$ is the surface gravity (see e.g., section 2 of ref.\cite{tp1}).
This covers the region outside the horizon ($N>0$) with the horizon mapping to the origin; disregarding the the region inaccessible
to the observers outside the horizon is equivalent to removing the origin from the $\tau-N$ plane. The contribution from the boundary of this region can be obtained by evaluating $A_{sur}$ on a surface infinitesimally close to the origin and taking the limit.
Consider a surface 
$N=\epsilon, 0<\tau<2\pi/\kappa$ and the full
range for the transverse coordinates; for $\epsilon \to 0$, this surface 
is infinitesimally away from the horizon in the Euclidean space described by Eq.(\ref{eext}) and has the unit normal $u^a=\kappa(0,1,0,0)$. Its contribution to  $A_{\rm sur}$ in Eq.(\ref{threedaction}) is the integral of $K=-\nabla_a u^a=-(\kappa/\epsilon)$ over the surface:
\begin{equation}
A_{\rm sur}=-\frac{1}{8\pi G}\int d^2 x_\perp \int_0^{2\pi/\kappa}d\tau 
\epsilon \left( \frac{\kappa}{\epsilon}\right) =-\frac{1}{4}\frac{\mathcal{A}_\perp}{G}
\label{seven}
\end{equation}  
which is (minus) one quarter of the transverse area $\mathcal{A}_\perp$ of the horizon, in units of $G$. This contribution is universal and --- because it is independent of $\epsilon$ --- the limit
$\epsilon\to0$ is trivial.  Since the surface contribution is due to removing the inaccessible region, it makes sense to identify $(-A_{\rm sur})$ with an entropy. (The sign in Eq.(\ref{threedaction}) is chosen with $G>0$ since we expect --- in the Euclidean sector --- the relation $\exp(-A_{\rm Euclid})=\exp S$ to hold, where $S$ is the entropy.)

The $A_{\rm sur}$, expressed as an integral over the extrinsic curvature term is closely related to the negative of the surface term of the Einstein-Hilbert action:
\begin{equation}
A_{sur,EH}=
\frac{1}{16\pi G}\int_{\mathcal{V}} d^4 x \, 
\partial_c[\sqrt{-g}Q_a^{\phantom{a}bcd}\Gamma^a_{bd}]=
\frac{1}{16\pi G}\int_{\partial\mathcal{V}} d^3 x \, 
\sqrt{-g}n_cQ_a^{\phantom{a}bcd}\Gamma^a_{bd}
\label{actfunc}
\end{equation}
where $Q_a{}^{bcd}=(1/2)(-\delta^c_ag^{bd}+\delta^d_ag^{bc})$. They agree on a specific gauge and their variations always match when metric is fixed on the surface. All the previous comments related to entropy continue to hold with this surface term and we shall hereafter use this term.

Given that $A_{sur}$ is related to the entropy, its variation has direct thermodynamic significance. To obtain the dynamics of spacetime continuum,
we will take the      
 total action $A_{tot}$ for matter plus gravity to be the sum of $A_{sur}+A_{matter}[\phi_i,g]$ where $A_{matter}[\phi_i,g]$ is the standard matter action  in a spacetime with metric
$g_{ab}$. The $\phi_i$ denotes some generic matter degrees of freedom; varying $\phi_i$
will lead to standard equations of motion for matter in a background metric and these equations will also ensure that the energy momentum tensor
of matter $T^a_b$ satisfies $\nabla_aT^a_b=0$.

We will now prove the key result of this paper: Einstein's equations arise 
from the demand that $A_{tot}=A_{sur}+A_{matter}$  should be invariant under virtual displacements of the horizon normal to itself. 
The physical interpretation is discussed after the mathematical derivation.

Let $\mathcal{V}$ be a region of spacetime such that part of the boundary of the spacetime $\partial\mathcal{V}$
is made up of the horizon $\mathcal{H}$. (For example, in the Schwarzschild metric we can take $\mathcal{V}$ to be bounded by the surfaces $t=t_1,t=t_2,r=2M,r=R>2M$. Whenever necessary, we will approach the horizon as a limit of a sequence of timelike surfaces like
$r=2M+\epsilon$ with $\epsilon\to0$.)
Consider
an infinitesimal coordinate transformation
$x^a\to \bar{x}^a=x^a+\xi^a(x)$, where $\xi^a(x)$ is nonzero \textit{only} on the horizon and is \textit{in the direction of the normal} to the horizon (which makes it a null vector). This transformation induces a (virtual) displacement of horizon normal to itself,
leaving the other bits of the boundary intact. The metric changes by
$\delta g^{ab}=\nabla^a\xi^b+\nabla^b\xi^a$ and the matter action changes by
\begin{equation}
\delta A_{matt}=-\frac{1}{2}\int_\mathcal{V}d^4x\sqrt{-g}T_{ab}\delta g^{ab}
=-\int_\mathcal{V}d^4x\sqrt{-g}\nabla_a(T^a_b\xi^b)
\label{delmat}
\end{equation}
where we have used the fact that $\nabla_a T^a_b=0$, which arises from the  equations of motion for the matter. Next, we need the explicit form of   $(\delta A_{sur}/\delta g^{ab})$ under infinitesimal coordinate transformations, which is given by:
\begin{equation}
\delta A_{\rm sur} = \frac{1}{8\pi  G} \int_{\mathcal{V}} d^4 x\,  \sqrt{-g}\, \nabla_a (R^a_b \xi^b)
\label{basic}
\end{equation}
We shall provide a quick derivation of Eq.(\ref{basic}) since  this result is not found in standard text books. This can be obtained directly by varying Eq.(\ref{threedaction}) but a cleverer procedure is the following: Recall that,
$A_{sur}$ in Eq.(\ref{threedaction}) is the usual extrinsic curvature term which is added to Hilbert action,
in order to cancel the variation in the term involving the second derivatives of the metric
(see e.g., Appendix A of ref.\cite{tp1}). Hence it follows that  
the variation of $(-A_{sur})$ is the same as that of the second derivative term in Hilbert action, mentioned (and then
ignored) in standard textbooks while deriving the Einstein's equations. Therefore, we have the result:
\begin{equation}
\delta (-A_{\rm sur})= \frac{1}{16\pi  G} \int_{\mathcal{V}} d^4 x\,  \sqrt{-g}g^{ab}\delta R_{ab}
\label{delasur}
\end{equation}
We will now show that under $x^a\to \bar{x}^a=x^a+\xi^a$,  the integrand in Eq.(\ref{delasur}) is $\sqrt{-g} g^{ab}\delta R_{ab}=-2\sqrt{-g}\nabla^a(R_{ab}\xi^b)$,
thereby establishing Eq.(\ref{basic}). To do this, note that
\begin{equation}
\delta [\sqrt{-g}R]=-\sqrt{-g}\nabla_a(R\xi^a)=\sqrt{-g}[G_{ab}\delta g^{ab}+g^{ab}\delta R_{ab}]
\label{varyR}
\end{equation}
The  first equality follows from the fact that
the \textit{local}
functional variation, $\delta [\sqrt{-g}Q(x)]$, of any scalar density $\sqrt{-g}Q(x)$ made from a generally covariant scalar $Q(x)$,
is  $\delta [\sqrt{-g}Q(x)]=-\sqrt{-g}\nabla_a(Q\xi^a)$. The second equality in Eq.(\ref{varyR}) is a standard text book result for $\delta [\sqrt{-g}R]$.
Using $\delta g^{ab}=\nabla^a\xi^b+\nabla^b\xi^a$, and $\nabla^aG_{ab}=0$
we now get:
\begin{eqnarray}
-\sqrt{-g}\nabla_a(R\xi^a)&=&2\sqrt{-g}\nabla^a[(R_{ab}-\frac{1}{2}g_{ab}R)\xi^b]+\sqrt{-g}g^{ab}\delta R_{ab}\\
&=&2\sqrt{-g}\nabla^a(R_{ab}\xi^b)- \sqrt{-g}\nabla_a(R\xi^a) + \sqrt{-g} g^{ab}\delta R_{ab}\nonumber
\end{eqnarray}
which immediately leads to the result $\sqrt{-g} g^{ab}\delta R_{ab}=-2\sqrt{-g}\nabla^a(R_{ab}\xi^b)$ 
completing the proof of Eq.(\ref{basic}). 

The rest is straightforward.
The integration of the divergences in Eqs.(\ref{delmat}),(\ref{basic}) leads to surface terms which contribute only on the horizon, since $\xi^a$ is nonzero only on the horizon.
Further, since $\xi^a$ is in the direction of the normal,
the demand $0=\delta A_{tot}=\delta A_{\rm sur} + \delta A_{\rm matter}$ leads to the result $(R^a_b-8\pi GT^a_b)\xi^b\xi_a=0$. Using the fact that $\xi^a$ is arbitrary \textit{except for being a null vector}, this requires  $R^a_b-8\pi GT^a_b=F(g)\delta^a_b$, where $F$ is an arbitrary function of the metric. Finally, since $\nabla_a T^a_b=0$ identically, $R^a_b-F(g)\delta^a_b$ must have  identical zero divergence; so $F$ must have the form
$F=(1/2)R+\Lambda$ where $R$ is the scalar curvature and $\Lambda$ is an undetermined (alas!) cosmological constant.
The resulting equation is
\begin{equation}
R^a_b-(1/2)R\delta^a_b+\Lambda\delta^a_b=8\pi GT^a_b
\end{equation}
which is identical to Einstein's equation. \textit{Nowhere did we need the bulk term in Einstein's action!} And the dynamics is independent of $u^i$ as it should.

We stress that this is a  totally new, \textit{self-contained}, perspective on gravity.  In this approach, the action functional for the continuum spacetime is
\begin{equation}
A_{tot}=A_{sur}+A_{matter}=\frac{1}{16\pi G}\int_{\partial\mathcal{V}} d^3 x \, 
\sqrt{-g}n_cQ_a^{\phantom{a}bcd}\Gamma^a_{bd}
+\int_{\mathcal{V}} d^4x \, \sqrt{-g}L_{matter}
\end{equation}
in which matter lives in the bulk $\mathcal{V}$ while the gravity contributes on the boundary ${\partial\mathcal{V}}$. When the boundary
has a part which acts as a horizon for a class of observers, 
we demand that the action should be invariant under virtual displacements of this horizon. This leads to Einstein's theory. Since $A_{sur}$ is related to the entropy,
 its variation, when the horizon is moved infinitesimally, is equivalent to the change in the entropy $dS$ due to virtual work. The variation
 of the matter term contributes the $PdV$ and $dE$ terms and the entire variational principle is equivalent to the thermodynamic identity $TdS=dE+PdV$ applied to the changes when a horizon undergoes a virtual displacement. In the case of spherically symmetric spacetimes, for example, it can be
\textit{explicitly} demonstrated \cite{ss} that the Einstein's equations follow from the thermodynamic
identity applied to horizon displacements. Since the current observations on dark energy is consistent [see e.g.,\cite{deobs}] with an
asymptotically deSitter universe with a horizon, this result will have implications for the explanation of cosmological constant [see \cite{ccpossible} for some possibilities].

The result also shows that Einstein's theory has an intrinsic holography. The standard description is in terms of $L_{bulk}$ and we have now shown that it has a dual description in terms of $L_{sur}$. It was noticed earlier \cite{tp1} that there is a remarkable relation between these two
terms 
\begin{equation}
\sqrt{-g}L_{sur}=-\partial_a\left(g_{ik}\frac{\partial \sqrt{-g}L_{bulk}}{\partial(\partial_a g_{ik})}      \right)
\end{equation} 
which has no explanation in standard approach.
The current analysis shows that the horizon entropy and resulting thermodynamics for local Rindler observers (based on $L_{sur}$)
leads to the same dynamics as that based on $L_{bulk}$, showing their interdependence.

Given the true microscopic degrees of freedom of spacetime (say, $q_i$) and an action
$A_{micro}$ describing them, the integration of $\exp(-A_{micro})$ over $q_i$, should lead to our $\exp(-A_{sur})$ as well as to the metric tensor, which is a macroscopic concept in the continuum limit,
analogous to, say,  the density field of a solid. (This approach has a long history\cite{sakharov}
but our result gives it a different, precise and elegant characterization). In the variation  $x^a\to \bar{x}^a=x^a+\xi^a$ the
$\xi^a(x)$ is similar to the displacement vector used, for example,
in the study of elastic solids. 
The true degrees of freedom are some unknown `atoms of spacetime' but in the continuum limit,
the displacement $x^a\to \bar{x}^a=x^a+\xi^a(x)$ captures the relevant dynamics,  just like 
in the study of elastic properties of the continuum solid. Further, it can be shown that the horizons in the spacetime are  similar to defects in the solid so that their displacement costs entropy. \footnote{There is subtlety here: As far as gravity is concerned, this can be thought of as  
 virtual displacements of this horizon normal
 to itself. But note that in the matter Lagrangian we are only varying $\delta g^{ab}=\nabla^a\xi^b+\nabla^b\xi^a$ and \textit{not} the matter fields $\phi$. In the case of a genuine active coordinate transformation, even matter fields will change, which is \textit{not} the case we are considering. We merely demand $\delta A_{tot}=0$ for a particular type of variation $\delta g_{ab}$ keeping everything else fixed.}

 Our demand that accelerated observers with horizons should be able to do consistent physics, with variables accessible to them,
turns out to be as powerful in determining the \textit{dynamics} of gravity, as the principle of equivalence (applied to inertial observers) was in determining the
\textit{kinematics} of gravity.

%\bibliography{forbhentropy}

\end{document}